# Power-over-fiber and distributed acoustic sensing hybridization in single fiber channel

Jing Zhang[1,2], Yao Chen[1], Tianye Huang[1,3,4,5*], Kaifu Chen[1], Hanlin Guo[1], Yongkang Huang[1], Lu Guo[1], Liangming Xiong[6], and Perry Ping Shum[7]

[1]School of Mechanical Engineering and Electronic Information, China University of Geosciences, Wuhan 430074, P.R. China

[2]School of Future Technology, China University of Geosciences, Wuhan 430074, P.R. China

[3]Wuhan National Laboratory for Optoelectronics, Hubei, P.R. China

[4]SINOPEC Key Laboratory of Geophysics, Nanjing, P.R. China

[5]Shenzhen Research Institute of China University of Geosciences, Shenzhen, P.R. China

[6]Yangtze optical fiber and cable Joint Stock Limited Company, Wuhan, Hubei, P.R. China

[7]Department of Electronic and Electrical Engineering, South University of Science and Technology, Shenzhen 518055, P.R. China

* Corresponding author: Tianye Huang (E-mail: huangty@cug.edu.cn)

## Abstract

The efficient and independent operation of power-over-fiber (PoF) and distributed acoustic sensing (DAS) has been demonstrated using standard single-mode fiber (SSMF). A transmission optical power efficiency (OPTE) of 6.67% was achieved over an 11.8 km fiber link, supporting both power delivery and distributed optical fiber sensing (DOFS). To minimize cross-talk, the system separates the power and sensing channels by a 40 THz bandwidth. In the experiment, the power and sensing light wavelengths are 1064 nm (continuous) and 1550 nm (pulsed), respectively. As the transmitted optical power increased from 0 W to 2.13 W, the DAS system successfully localized vibration sources and reconstructed phase information, confirming its ability to operate under high optical power. The reported scheme verifies the possibility of constructing the sensing-energy hybrid network based on conventional optical fiber with the advantages of flexibility and low cost.

# 1. Introduction

Power-over-Fiber (PoF) technology enables the transmission of electrical power through optical fibers, leveraging the high efficiency and safety of optical systems. PoF eliminates the need for copper cables, reducing electromagnetic interference, power losses, and weight, while enhancing the flexibility and reliability of power transmission, particularly in remote, hazardous, or difficult-to-reach locations. This technology has a wide range of applications, including remote sensor networks [1], 5G remote antenna units [2], offshore platforms [3], and underground monitoring systems [4], where traditional power delivery would be costly or impractical.

The development of PoF technology dates back to the late 1970s [5]. Integrating with radio-over-fiber (RoF) systems, 250 mW optical power is delivered over 300 m multimode fiber [6]. In order to increase the power level, double-cladding fibers are employed to transmit optical power up to 60 W [7]. In multi-core fiber (MCF), the power light and signal light at the same 1.55 μm band are transmitted with different separate cores. However, the total optical power arising in the seven-core fiber is generally smaller than 1 W, due to the cross-talk and nonlinearity [8-9]. PoF systems supporting 5G cellular networks and internet of things (IoT) applications based on standard single-mode fibers (SSMF) is proposed to deliver 2 W of optical power over distances ranging from 100 m to 1 km [10]. Yang et al. further illustrated the efficiency of PoF by co-transmitting 10 W of power light with a 5G NR signal over a 1-km SSMF link. Their approach effectively mitigated nonlinear effects through precise linewidth manipulation and wavelength spacing adjustments, confirming the capability of PoF systems to support high-speed data communication [11]. To increase the damage threshold, anti-resonant hollow-core fibers (AR-HCF) are good candidates for power delivery. The experiments demonstrated near-diffraction-limited beam quality and no fiber damage during 300 W high-power delivery, providing valuable insights into the practical limitations and optimization strategies for high-power PoF systems [12].

Besides power delivery, optical fiber is also a promising platform for distributed sensing due to the advantages of long distance, high sensitivity, and high spatial resolution, which made optical fiber indispensable in fields such as seismic monitoring [13], structural health assessment [14], vertical seismic profile [15-16], and temperature sensing, etc. However, in remote or hazardous environments, such as deep-sea oil platforms, underground tunnels, or vast pipelines, traditional electrical power infrastructure is either impractical or too costly to deploy. While, in these scenarios, besides distributed optical fiber sensors (DOFS), multiple modalities of monitoring are required to reduce the false alarm rate [17-19]. The simultaneous implementation of DOFS and other electronic sensors is necessary. Therefore, power delivery becomes highly significant to ensure energy supply for other sensing systems.

The integration of PoF and DOFS in a single optical fiber offers a unique advantage by combining the benefits of both technologies into a single, unified system. This combination enables the simultaneous delivery of power for other kinds of sensors and real-time fiber sensing along the same fiber network. As mentioned above, the performance of the PoF system is highly dependent on the fiber type. Double-cladding fiber and AR-HCF allow for high power transmission but they are not widely deployed. Particularly, AR-HCF is not the best choice for DOFS due to the low backscattering strength. In contrast, due to the enormous demand for the IP service, SSMF is still the most popular and widely implemented fiber type. The damage threshold of the SSMF permits more than 10 W optical power delivery [20] makes it a good candidate for simultaneous power delivering and distributed fiber sensing.

In this paper, a cost-effective approach to simultaneously realize the PoF and distributed acoustic sensing (DAS) is experimentally verified. Specifically, the power channel and the sensing channel are arranged in the 1.0 μm and 1.55 μm bands, respectively, with a bandwidth separation of more than 40 THz to avoid nonlinear power transfer in SSMF. We show that 2.13 W power light co-transmits with sensing probe light over SSMF length of 11.8 km, where the collected power of the power light is 142 mW and the sensing performances are well-maintained. The synergy between PoF and DAS not only simplifies the deployment of multi-parameter sensor networks in remote or hard-to-access areas but also enhances the system's scalability, flexibility, and cost-effectiveness.

## 2. Experiments

The experimental setup for the proposed PoF-DAS system is shown in Fig.1. The light source for the DAS system is a narrow linewidth laser (NLL) at 1550.12 nm with a linewidth of 2 kHz. In the upper branch, the continuous-wave (CW) light is modulated by an acoustic-optic modulator (AOM) into pulsed light with a pulse width and repetition rate of 100 ns and 8 kHz, respectively. The pulsed light was amplified by an erbium-doped fiber amplifier (EDFA) to boost the peak power which is essential for extending the sensing range and improving the signal-to-noise ratio (SNR). A circulator was used to send detection pulses to the fiber under test (FUT) with a total length of 11.8 km. A piezoelectric (PZT) transducer is placed approximately 10 km from the fiber to generate vibration events, with the frequency controlled by a function generator. The lower-branch light is used as the local oscillator (LO) light for DAS, enabling self-heterodyne coherent detection. A balanced photon detector (BPD) received the coherent signal and converted it into an electrical signal, which is then observed using an oscilloscope (OSC) to collect the output. The obtained digital data was transported to a computer for further processing.

For PoF, a wavelength division multiplexer (WDM) is inserted between the circulator and the sensing fiber to achieve a perfect integration of the PoF function with the DAS system. At the same time, a continuous-wave laser at 1064 nm served as the PoF source. The PoF-DAS hybrid system share the same optical fiber. At the end of the fiber link, another WDM separates the power band and the sensing band from the fiber. The power of power light is measured by using an optical power meter (Thorlabs, S415C).

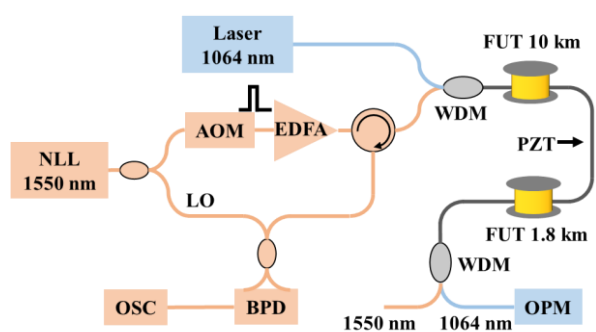

Fig. 1 The experimental setup for the co-transmission systems.

## 3. Results and discussions

Figure 2 (a) presents the collected power of the power light both with or without DAS at the end of the 11.8 km SSMF. Fig. 2(b) shows the corresponding loss and optical power transmission efficiency (OPTE), which is defined as the ratio of collected optical power to the injection power of power light. When 2.13 W continuous power light is introduced into the 11.8 km SSMF, the maximum collected optical power is 142 mW, with a corresponding OPTE of about 6.67%. While this value may seem low, but it is consistent with the transmission loss of 1064 nm light in SSMF [21]. As shown in Fig. 2(b), the transmission loss over 11.8 km is about 1 dB/km at 1064 nm for ordinary SSMF. The relatively low OPTE is primarily due to the inherent attenuation of optical fibers at this wavelength, which is a significant challenge and an important factor that must be considered in long-distance optical power transmission systems. To improve higher power efficiency, shorter fiber lengths can be considered. However, this way would limit the sensing range of the DAS system, so there is a trade-off between OPTE and the sensing range. Fig. 2 also shows the power transmission performance of the PoF-DAS hybrid system. The results confirm that PoF operation is independent of DAS functionality, which is a key advantage of the hybrid system, especially for applications requiring multi-sensor networks.

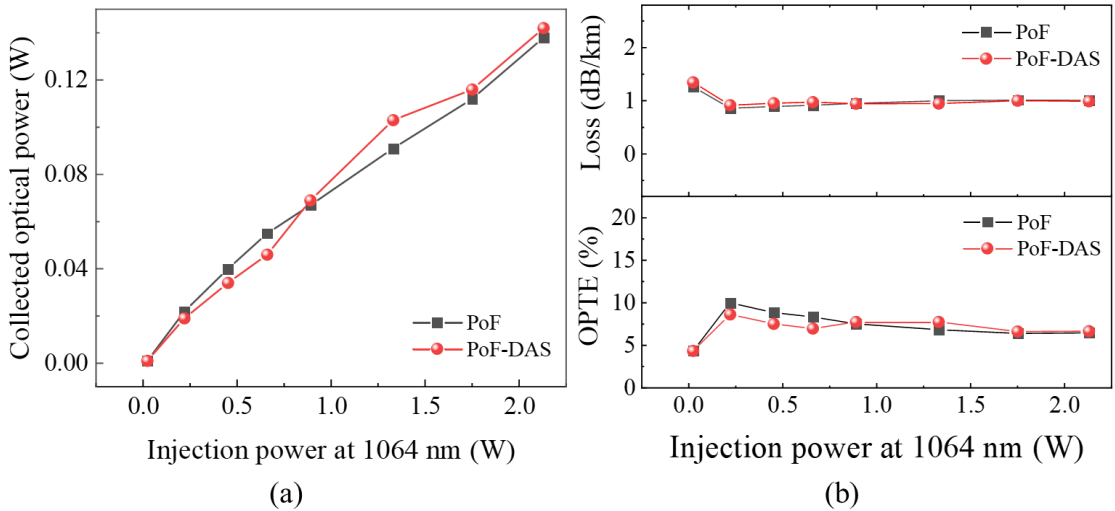

Fig. 2 (a) Collected power of power light, (b) corresponding loss and corresponding OPTE under different injection powers.

Figure 3 shows the vibration localization results of the DAS under different delivered power. The vibration source in the DAS system is set as a sinusoidal signal with a frequency of 500 Hz and an amplitude of 5 V. The vibrational localization results can be clearly seen that the vibration event is located at 10.18 km, which is consistent with the test fiber setup in the experiment. As the power of 1064 nm laser increases (0 W to 2.13 W), the noise in the vibrational localization results also increases, indicating greater jitter in the backward Rayleigh scattering signal. Although this effect is present, it is weak within the tested 1064 nm optical power range. Of course, some more advanced methods, such as wavelet transform [22], could enhance the SNR and improve localization accuracy compared to basic methods like moving differential method used. Overall, the vibrational localization results indicate that the setup two-channel system works properly.

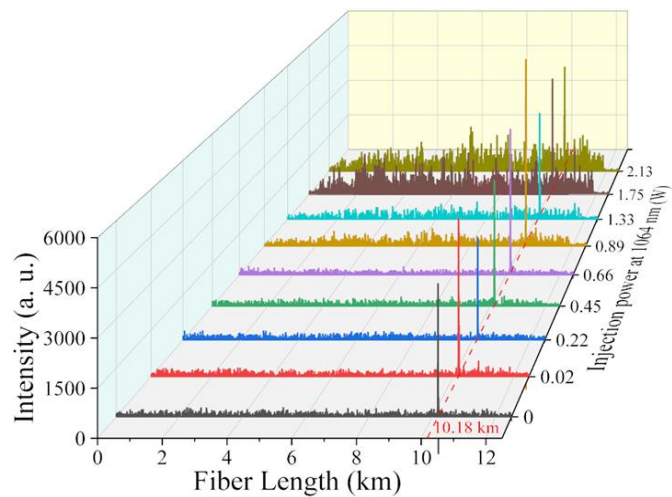

Fig. 3 The vibration localization results of DAS under different delivered power.

Once the DAS system identifies the location of the disturbance, another important work is to demodulate the external vibration source. Figure 4 shows the phase demodulation results of the DAS under different delivered power. The results indicate that the basic sensing capability of the system remains intact, although the data quality may be compromised and more advanced signal processing techniques are required to extract data with higher quality. To mitigate noise, complex signal processing and interference fading mitigation algorithms may be required to suppress the signal fluctuations [23, 24].

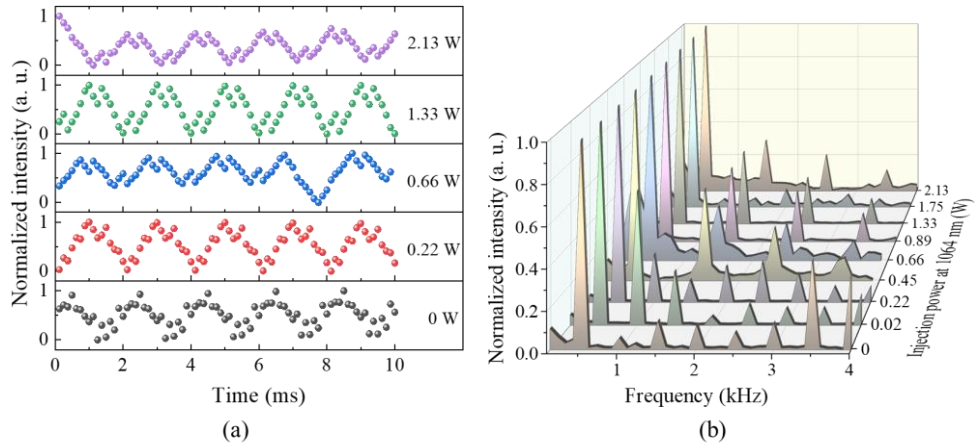

Fig. 4 (a) The demodulation results, and (b) the normalized spectral results of DAS under different delivered power.

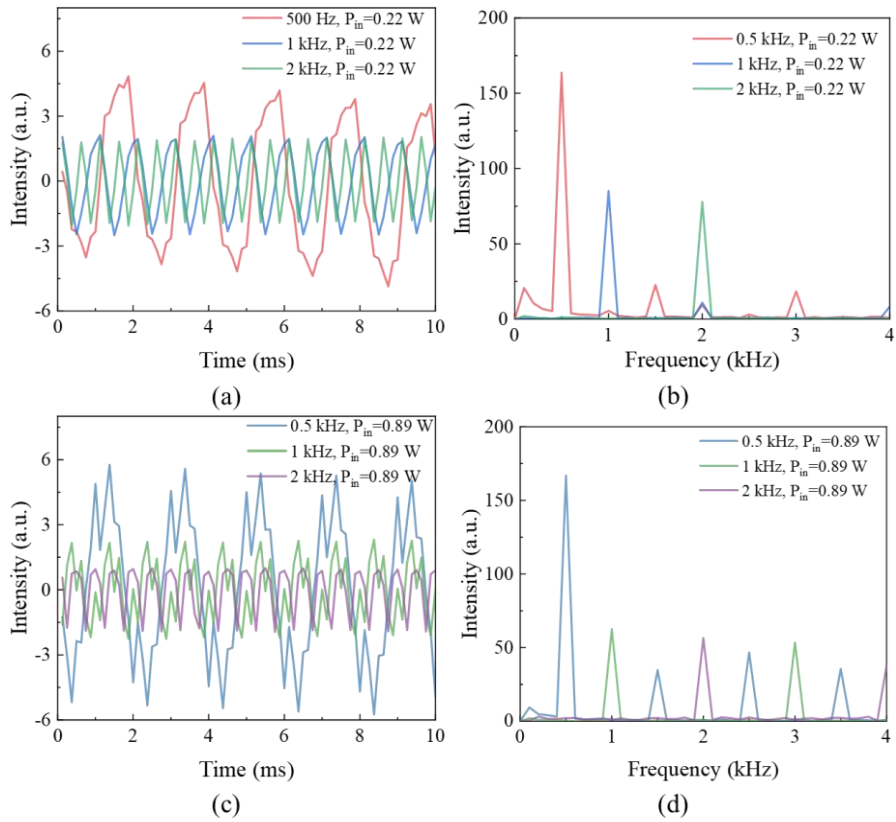

Fig. 5 The demodulation results and the frequency response of DAS at 1064 nm power of 0.22 W and 0.89 W

The frequency response of the DAS was examined under different injection powers, and different vibration frequencies (500 Hz~2 kHz) were set for the same amplitude of the PZT (5 V). Figs. 5 show that the DAS responds to different vibration frequencies, although high-order frequency noise is observed, which may be attributed to the simple arctangent demodulation method used. As shown in Fig. 4(b), the frequency response results without 1064 nm-light injection also have higher-order frequency noise. Note that this is not introduced by energy co-transmission, while appears in the original DAS system. Although the 1064 nm optical power increases, the corresponding fundamental capabilities for detecting different frequencies remain unaffected. Overall, these results confirm that presence of 1064 nm light in the fiber does not impact the performance of the DAS system, whether for vibration localization or phase demodulation.

To further evaluate the demodulation performance of the system, the power spectral density (PSD) of the vibration signal was calculated to assess the strength of the interference signal. Fig. 6(a) and (b) show the PSD at 100 Hz and 2 kHz, respectively. The calculated SNRs were found to be 28.39 dB and 34.98 dB, respectively. These high SNRs values indicate that the system is ability to effectively demodulate interfering signals, even in the presence of noise, which is critical for accurate vibration detection and localization.

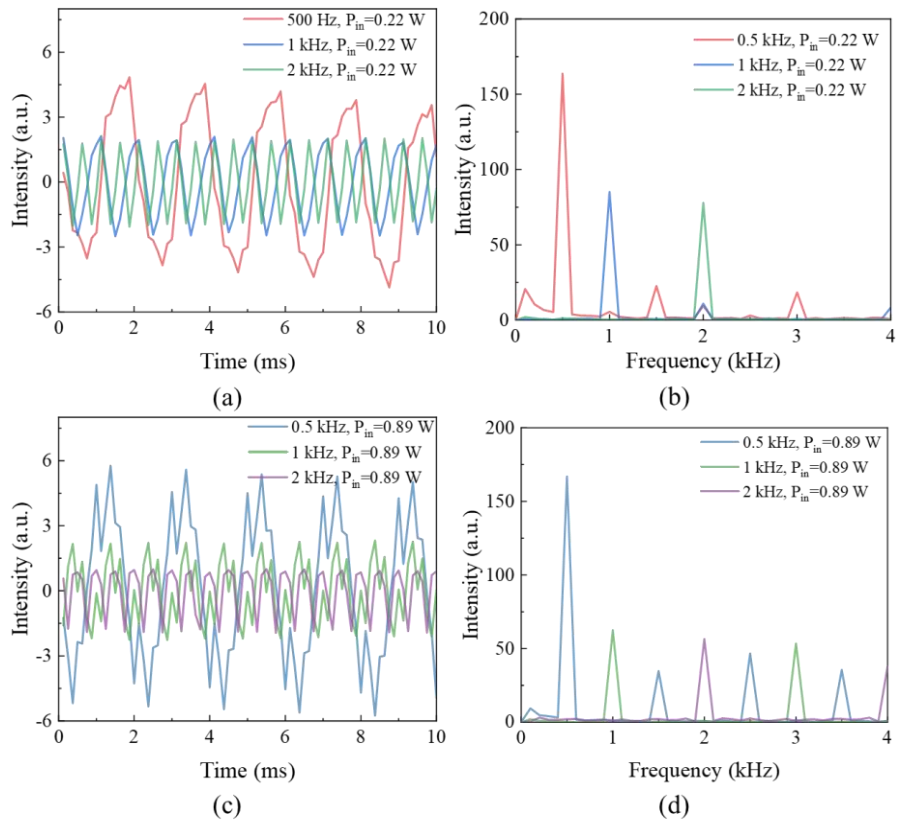

Fig. 6 The PSD results for dynamic measurements at (a) 100 Hz and (b) 2 kHz

## 4. Conclusion

In conclusion, we have demonstrated, for the first time, the co-transmission of 2.13 W power light and DAS signals over 11.8 km SSMF. As the 1064 nm optical power increases, the vibrational localization and frequency response for the DAS system was evaluated, confirming that DAS and PoF can be coexistence simultaneously on the same fiber with minimal impact on each other within the tested power range. Since the SSMF has been widely deployed with corresponding off-the-shelf components, the SSMF-based PoF-DAS hybrid system expected to form a new technological pathway for constructing multi-sensor network in the difficult-to-access areas.

**Funding**. This work is supported by the National Natural Science Foundation of China (42327803, 62275097, 62205313); Open Fund of SINOPEC Key Laboratory of Geophysics; Open Project Program of Wuhan National Laboratory for Optoelectronics (2023WNLOKF007); Technology Innovation Project of Hubei Province (2022BEC003); Key R&D Program of Hubei Province (2023BAB062); Major Science and Technology Projects of Wuhan (2023010302020030); Guangdong Basic and Applied Basic Research Foundation (2023A1515010965, 2024A1515010017); Postdoctoral Fellowship Program of CPSF (2023 M733301); Fundamental Research Funds for Central University China University of Geosciences (Wuhan) (CUGDCJ202204, G1323523064).